# LINE-CONSTRAINED GEOMETRIC SERVER PLACEMENT

Mugurel Ionuţ ANDREICA, Eliana-Dina TÎRŞA

Computer Science Department, Politehnica University of Bucharest, Bucharest, Romania
email: {mugurel.andreica, eliana.tirsa}@cs.pub.ro

**Abstract:** *In this paper we present new algorithmic solutions for several constrained geometric server placement problems. We consider the problems of computing the 1-center and obnoxious 1-center of a set of line segments, constrained to lie on a line segment, and the problem of computing the K-median of a set of points, constrained to lie on a line. The presented algorithms have applications in many types of distributed systems, as well as in various fields which make use of distributed systems for running some of their applications (like chemistry, metallurgy, physics, etc.).*

## 1. INTRODUCTION

Server placement is an important task in many types of distributed systems, e.g. (multimedia) content delivery networks (selecting the placement of the content delivery servers), data storage services (selecting the machines on which replicas should be placed), Clouds (selecting the machines on which a client application should be run, in order to meet the given constraints) and Grids (e.g. for running scientific simulations from various fields, like chemistry, metallurgy or physics). In this paper we focus on several constrained geometric server placement problems. Geometric constraints may occur in physical settings (e.g. the placement of *relay nodes* in wireless networks) or when the problem's parameters are modeled using a (multidimensional) metric space. Occasionally, such problems can be expressed as (k-) center or (k-)median problems in the metric space. In Section 2 we present new solutions for computing the 1-center and obnoxious 1-center of a set of line segments, constrained to lie on a line segment. In Section 3 we consider the problem of computing the K-median of a set of points, constrained to lie on line. In Section 4 we discuss related work and in Section 5 we conclude and present future work.

## 2. CENTERS ON A LINE FOR A SET OF LINE SEGMENTS

We consider $N$ line segments in the plane. We want to compute the largest empty circle and the smallest enclosing circle whose center is constrained to lie on a given line segment $(xa,ya)-(xb,yb)$. The distance we use is the $L_p$ norm $(1 \leq p < +\infty)$.

Note that the smallest enclosing circle is equivalent to finding a server location such that the distance from the server to the furthest point on one of the segments (resources) is minimum. The largest empty circle problem is equivalent to finding the location of an obnoxious server, i.e. a server for which the minimum distance from it to some point on one of the segments is as large as possible.

### 2.1. Covering Interval

The proposed solutions to the smallest enclosing circle and largest empty circle problems start by rotating the coordinate system such that $(xa,ya)$ is moved to $(0,0)$ and $(xb,yb)$ is moved to $(L,0)$, where $L$ is the length of the segment $(xa,ya)-(xb,yb)$. Given a value $R$, we define the covering interval $[u,v]$ of a segment $(x_1,y_1)-(x_2,y_2)$ as the interval of points on the $OX$ axis which are located at distance at most $R$ from the segment $(x_1,y_1)-(x_2,y_2)$. W.l.o.g., we will assume that $x_1 \leq x_2$. We need to consider several cases:

- case *1*: $y_1 \leq R$ and $y_2 \leq R$ and the segment $(x_1,y_1)-(x_2,y_2)$ does not intersect *OX*
- case *2*: $y_1 \leq R$ and $y_2 > R$ and the segment $(x_1,y_1)-(x_2,y_2)$ does not intersect *OX*
- case *3*: $y_1 > R$ and $y_2 \leq R$ and the segment $(x_1,y_1)-(x_2,y_2)$ does not intersect *OX*
- case *4*: $y_1 > R$ and $y_2 > R$ and the segment $(x_1,y_1)-(x_2,y_2)$ does not intersect *OX*
- case *5*: the segment $(x_1,y_1)-(x_2,y_2)$ intersects *OX*

In the first case, we will have $u = x_1 - (R^p - y_1^p)^{1/p}$ and $v = x_2 + (R^p - y_2^p)^{1/p}$. In case we cannot compute the expressions directly, we can binary search the values $u$ and $v$ with a given precision $\varepsilon$, such that: $u$=the smallest value on the interval $(-\infty, x_1]$ such that $distance((u,0), (x_1,y_1)) \leq R$ and $v$=the largest value on the interval $[x_2, +\infty)$ such that $distance((v,0), (x_2,y_2)) \leq R$. In case *2* we compute $u$ as in case *1*. Then, we need to find the point $(x_3, y_3)$ along the segment $(x_1,y_1)-(x_2,y_2)$ such that $y_3 = R$. We have $(x_3-x_1)/(x_2-x_1) = (R-y_1)/(y_2-y_1)$, thus obtaining $y_3$. Alternatively, $x_3$ can be binary searched (with a fixed precision $\varepsilon$) as the largest value in the interval $[x_1, x_2]$ for which $y_1 + (y_2-y_1) \cdot (x_3-x_1)/(x_2-x_1) \leq R$. We have $v = x_3$. In case *3* we compute $v$ as in case *1* and then we need to find the point $(x_3, y_3)$ along the segment $(x_1,y_1)-(x_2,y_2)$ such that $y_3 = R$. We can proceed as in the previous case. For the binary search solution, $x_3$ will be the smallest value on the interval $[x_1, x_2]$ for which $y_1 + (y_2-y_1) \cdot (x_3-x_1)/(x_2-x_1) \leq R$. We have $u = x_3$. In the $4^{th}$ case the interval $[u,v]$ will be empty (i.e. the covering interval is void). In this case we can set $u = +\infty$ and $v = -\infty$. In the $5^{th}$ case we will compute the intersection point $(xi, 0)$ where the segment $(x_1,y_1)-(x_2,y_2)$ intersects *OX*. Then, we will consider the segment to be composed of two disjoint segments which (conceptually) do not intersect *OX*: $(x_1,y_1)-(xi,0)$ and $(xi,0)-(x_2,y_2)$. We compute the covering intervals separately for each of these two segments (using one of the cases *1-4*) and then $[u,v]$ is defined as the union of the covering intervals of the two smaller segments.

### 2.2. Smallest Enclosing Circle

We will binary search the radius *Rmin* of the smallest enclosing circle of all the $N$ given segments (with a fixed precision $\varepsilon$) in the interval $[0, +\infty)$. The property we use is the following: if $R < Rmin$ then the intersection of the covering intervals of the $N$ segments and the interval *[0,L]* is void; if $R \geq Rmin$ then the intersection of the $N$ covering intervals and the interval *[0,L]* is non-void. Thus, *Rmin* is the smallest value for which the intersection of the covering intervals and the interval *[0,L]* is non-empty. The time complexity of this solutions is $O(N \cdot log(Rmin))$. We need only $O(N)$ time in order to compute all the covering intervals for a given value of $R$ and in order to decide if the intersection of the $N+1$ intervals is empty (their intersection is empty if the largest left endpoint of an interval is larger than the smallest right endpoint).

The *x* coordinate of the circle can be any of the points of the intersection interval obtained when $R=Rmin$ (normally, this interval should consist of just one point; however, since the binary search is performed with a fixed precision $\varepsilon$, the length of the intersection interval is bounded by a function of $\varepsilon$).

### 2.3. Largest Empty Circle

A solution based on binary search is possible for this problem, too. The maximum radius *Rmax* of an empty circle has the following property: for $R<Rmax$ the union of the covering intervals of the $N$ segments covers the interval *[0,L]* completely; for $R \geq Rmax$ at least some part of *[0,L]* is not covered by any of the $N$ covering intervals. We need $O(N)$ time in order to compute the $N$ covering intervals and $O(N \cdot log(N))$ in order to compute their union (because the intervals need to be sorted). Then, in $O(N)$ time, we can check if the covering intervals' union fully covers *[0,L]* or not. Thus, this solution has a time complexity of $O(N \cdot log(N) \cdot log(Rmax))$.

This problem also has a different solution, based on computing the lower envelope of the distance functions of the $N$ line segments. The method implements a lower envelope computation algorithm based on the divide-and-conquer solution presented in [1]. The lower envelope consists of a set of disjoint intervals (except possibly for their endpoints) *[a(i),b(i)]* whose union is equal to *[0,L]*. For each interval *[a(i),b(i)]*, the segment *seg(i)* which is closest among all the other segments to every point *(x,0)* ($a(i) \leq x \leq b(i)$) is also stored. A pseudocode of the lower envelope computation algorithm is given below:

**ComputeLowerEnvelope(S: set of segments)**
  **if** *|S|=1* **then {**
    Let *segm* be the only segment in *S*.
    Let $xmin_{segm}$ be the x coordinate of the point on the OX axis at which the segment *segm* is the closest to the interval *[0,L]*
    Construct a lower envelope consisting of two intervals: *[a(1)=0,b(1)=$xmin_{segm}$]* and *[a(2)=$xmin_{segm}$, b(2)=L]* (and *seg(1)=seg(2)=segm*).
  **} else {**
    Split *S* into two (disjoint) sets $S_1$ and $S_2$
    $LE_1$=ComputeLowerEnvelope($S_1$)
    $LE_2$=ComputeLowerEnvelope($S_2$)
    **return** *MergeLowerEnvelopes($LE_1$,$LE_2$)*
  **}**

In order to compute $xmin_{segm}$ we have several options. The most general one is to use a binary search approach on the interval *[0,L]*. We have $x<xmin_{segm}$ if *distance((x,0), segm)>distance((x+$\varepsilon$,0), segm)* and $x \geq xmin_{segmn}$ if *distance((x,0), segm)$\leq$distance((x+$\varepsilon$,0), segm)*. This approach basically approximates the first derivative of the distance function. The distance function is unimodal, i.e. it decreases up to the point $xmin_{segm}$ and then it increases. This means that up to $xmin_{segm}$, the first derivative is negative, after which it becomes positive.

Another solution of computing $xmin_{segm}$ is based on a more geometric approach. If *segm* intersects *[0,L]*, then $xmin_{segm}$ is the *x* coordinate of the intersection point. Otherwise, $xmin_{segm}$ is either *0*, *L*, $x_1$ or $x_2$ (where *segm=$(x_1,y_1)$-$(x_2,y_2)$*). $x_1$ and $x_2$ are considered as candidates only if they are located in the interval *[0,L]*. Then, by computing the distance from each of the candidate points to *segm* and choosing the minimum value we can easily find $xmin_{segm}$.

The merging of two lower envelopes $LE_1$ and $LE_2$ is performed as follows. We sort all the interval endpoints of both lower envelopes (removing duplicates). Note the since the endpoints of each of the two lower envelopes $LE_1$ and $LE_2$ are already sorted, we can implement this step in $O(|LE_1|+|LE_2|)$ time (by simply using the linear merge step of the merge sort algorithm). Then, for every two consecutive endpoints *u* and *v* in the sorted order, there is one segment $seg_1$ closest to the interval *[u,v]* in $LE_1$ and one segment $seg_2$ closest to *[u,v]* in $LE_2$ ($seg_1$ corresponds to the interval *[a(i),b(i)]* of $LE_1$ which includes *[u,v]*; $seg_2$ corresponds to the interval of $LE_2$ which includes *[u,v]*). Then, we have the following cases:

1) $seg_1$ is closer to both *(u,0)* and *(v,0)* than $seg_2$ : then the merged lower envelope contains the segment *[u,v]* and $seg_1$ is associated to it

2) $seg_2$ is closer to both *(u,0)* and *(v,0)* than $seg_1$ : then the merged lower envelope contains the segment *[u,v]* and $seg_2$ is associated to it

3) $seg_1$ is closer to *u* than $seg_2$ and $seg_2$ is closer to *v* than $seg_1$ : then we compute the point *w* ($u \leq w \leq v$) such that $seg_1$ and $seg_2$ are located at equal distance from *(w,0)*: the merged lower envelope will contain the intervals *[u,w]* (with $seg_1$ associated to it) and *[w,v]* (with $seg_2$ associated to it)

4) $seg_2$ is closer to *(u,0)* than $seg_1$ and $seg_1$ is closer to *(v,0)* than $seg_2$ : then we compute the point *(w,0)* ($u \leq w \leq v$) such that $seg_1$ and $seg_2$ are located at equal distance from *(w,0)*: the merged lower envelope will contain the intervals *[u,w]* (with $seg_2$ associated to it) and *[w,v]* (with $seg_1$ associated to it)

Note that the intersection between any line segment (*[0,L]* in particular) and the bisector of two line segments has at most three intersection points. This is the same as saying that the distance functions (on the interval *[0,L]*) of two line segments $seg_1$ and $seg_2$ intersect in at most three points. Let's assume w.l.o.g. that $xmin_{seg1} \leq xmin_{seg2}$. Then the first intersection point may only be located before $xmin_{seg1}$, the second intersection point between $xmin_{seg1}$ and $xmin_{seg2}$, and the third intersection point after $xmin_{seg2}$.

The lower envelopes $LE_1$ and $LE_2$ contain among their endpoints the points $xmin_{segm}$ where *segm* is one of the segments which is closest to some of the envelopes' intervals. Thus, when we reach cases 3 or 4 (presented above), the segments $seg_1$ and $seg_2$ intersect in exactly one point *w*. This point can be binary searched on the interval *[u,v]*. We have $x \leq w$ if *distance((x,0), $seg_1$)$\leq$distance((x,0), $seg_2$)* (in case *3*) or if *distance((x,0),$seg_2$)$\leq$distance((x,0), $seg_1$)* (in case *4*). We have $x>w$ if *distance((x,0), $seg_1$)>distance((x,0), $seg_2$)* (in case *3*) or if *distance((x,0), $seg_2$)>distance((x,0), $seg_1$)* (in case *4*).

In the end, the merged lower envelope is compacted. The rule we use (repeatedly) is the following: if two consecutive intervals of the merged lower envelope *[a(i),b(i)]* and *[a(i+1),b(i+1)]* (where *b(i)=a(i+1)*) have the same segment associated to them *seg(i)=seg(i+1)* and *b(i)* is not the point at which the distance function of the segment *seg(i)* attains the minimum value on the interval *[0,L]* (i.e. $xmin_{seg(i)}$), then the two intervals are merged into a single one: *[a(i),b(i+1)]* whose associated segment is *seg(i)*. This algorithm can be implemented in *O(N)* time, by using a single traversal of the merged lower envelope.

Notice that we make sure that the fractions where minimum distances are attained for segments associated to intervals of the lower envelope are always kept as endpoints (and not located strictly within the interior of some interval). This way we can guarantee that on every interval *[u,v]* of the merged lower envelope of two lower envelopes the distance functions of the two considered segments intersect in at most one point (we used this property in the cases 1-4 above).

If we choose $|S_1|=|S|-1$ and $|S_2|=1$ then the time complexity of the presented algorithm is $O(N^2 \cdot log(w))$. If we choose $|S_1|$ approximately equal to $|S_2|$ (*S* is split into two approximately equal sets) then the time complexity is $O(N \cdot log(N) \cdot log(w))$. The *log(w)* factor is given by the computation of the intersection of the distance functions of two segments on a fixed interval *[u,v]* (using binary search). If this step can be computed analytically in *O(1)* time, then the *log(w)* factor can be dropped. The center of the largest empty circle is one of the endpoints of the computed lower envelope. The radius of the largest empty circle with center at an endpoint *a(i)* or *b(i)* is the distance from *a(i)* (or *b(i)*) to *seg(i)*. Fig. 1 shows an instance of the problem.

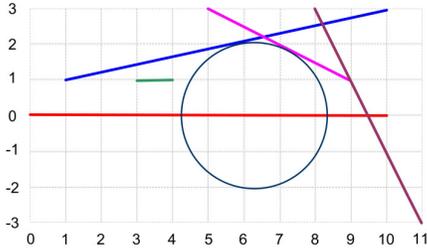

**Fig. 1. Largest empty circle with the center in the interval *[0,L=10]* on the *OX* axis, among a set of *N=4* line segments.**

For testing purposes, we implemented the previous algorithm with two versions (version *1*: $|S_1|=|S|/2$ and version *2*: $|S_1|=|S|-1$) and we evaluated them on random sets of *N=100,000* line segments. We found the running times to be quite similar, although the theoretical time complexities are very different. This is because, in most cases, the lower envelopes contained just a small number of intervals (e.g. around *O(log(N))* such intervals), making the two solutions behave almost identically.

## 3. K-MEDIANS ON A LINE FOR A SET OF POINTS

This problem generalizes the smallest enclosing circle problem from the previous section, but replaces the *N* line segments by *N* points. We need to place *K* circles with their centers anywhere on the *OX* axis, such that each of the *N* points is contained in at least one circle (inside or on its border) and an aggregate (e.g. sum) of the $q^{th}$ ($q \geq 1$) power of the radii of the *K* circles needs to be minimized. Note that we still consider the generalized version of the circle (in which the $L_p$ norm is used). The major observation for solving this problem is the following. Let's consider the points sorted in increasing order of their *x* coordinate: $x(1) \leq x(2) \leq \ldots \leq x(N)$. Then the optimal solution can be obtained as follows. We can split the interval *[1,N]* into *K* disjoint intervals *[l(i),r(i)]* ($l(i) \leq r(i)$; $1 \leq i \leq K$) such that *l(\*)* and *r(\*)* are integers between *1* and *N* and every integer number between *1* and *N* belongs to exactly one of these *K* intervals. Then, we compute the *K* smallest enclosing circles with the center on the *OX* axis for each set of points *x(l(i))*, *x(l(i)+1)*, …, *x(r(i))*. A dynamic programming algorithm goes as follows. We compute *OPT(i)*=the optimal solution if we consider only the points *x(1)*, …, *x(i)*. We have *OPT(0)=0* and $OPT(1 \leq i \leq N)=min\{agg(OPT(j-1), Rmin(j,i)^q) \mid 1 \leq j \leq i$ and *Rmin(j,i)* is the minimum radius of a circle with the center on the *OX* axis which contains all the points *(x(j), y(j))*, *(x(j+1), y(j+1))*, …, *(x(i-1), y(i-1))*, *(x(i), y(i))*}. We can compute *Rmin(j,i)* in $O((i-j+1) \cdot log(Rmin(j,i)))$ time, using the algorithm presented earlier for the smallest enclosing circle (where we consider *L=+∞*). This would lead to an $O(N^3 \cdot log(Rmin))$ time algorithm. In order to improve the time complexity of the solution we need the following function: *C(i,j)=(xc(i,j),R(i,j))* returns the *x* coordinate of the circle and the radius of the smallest enclosing circle (with the center on the OX axis) which fully contains both the points *(x(i),y(i))* and *(x(j),y(j))*. *xc(i,j)* can be easily computed as the intersection of the bisector of the points *(x(i),y(i))* and *(x(j),y(j))* with OX. If *i=j* the *xc(i,i)=x(i)*. The centers of the *K* circles to be placed must be among the points *xc(i,j)* ($1 \leq i \leq j \leq N$).

We will maintain a list *L(j)* for each position *j* (all the lists are initially empty). Then, we will consider all the pairs *(i,j)* ($1 \leq i \leq j \leq N$) and compute *xc(i,j)* and *R(i,j)*. We initialize *pleft(i,j)=i*. Then, while *pleft(i,j)>1* and the point *(x(pleft(i,j)-1),y(pleft(i,j)-1))* is at a distance at most equal to *R(i,j)* from *xc(i,j)* then we decrement *pleft(i,j)* by *1*. Similarly, we initialize *pright(i,j)=i* and then, while *pright(i,j)<N* and the point *(x(pright(i,j)+1),y(pright(i,j)+1))* is located at a distance at most equal to *R(i,j)* from *xc(i,j)* we increment *pright(i,j)* by *1*. We add the pair *(left=pleft(i,j), radius=R(i,j))* to the list *L(pright(i,j))*. This stage takes $O(N^3)$ time overall. Then, we will compute *OPT(i)* as follows: *OPT(0)=0* and $OPT(1 \leq i \leq N)=min\{agg(OPT(left-1), radius) \mid (left, radius)$ belongs to *L(i)*}. Note that *(i,y(i))* always belongs to *L(i)* ($1 \leq i \leq N$). The dynamic programming stage takes only $O(N^2)$ time, as there are only $O(N^2)$ pairs overall in all the lists *L(i)* ($1 \leq i \leq N$). In this case, the overall time complexity is dominated by the stage which computes the lists *L(\*)*.

A faster way of computing these lists is the following. We will first consider all the pairs *(i,j)* ($1 \leq i \leq j \leq N$), compute *xc(i,j)* and *R(i,j)*, and then we sort the values *xc(i,j)* in ascending order. This stage takes so far only $O(N^2 \cdot log(N))$ time. Then, we sweep the plane with a vertical line form left to right (from -∞ to +∞). During the sweep we will maintain a list of points located to the left of the sweep line sorted in increasing order of their distance to the point *(xs,0)*, where *xs* is the current *x* coordinate of the sweep line. During the sweep we encounter three types of events:

- type *1*: two points *(x(i),y(i))* and *(x(j),y(j))* which are adjacent in the distance ordering swap their order in the distance-based ordering
- type *2*: a new point *(x(i),y(i))* is encountered
- type *3*: a circle center *(xc(i,j),0)* is encountered

If multiple events occur simultaneously (at the same *x* coordinate), then we prefer the events in increasing order of their type (first type *1* events, then type *2* and then type *3*).

When a type *3* event occurs (the circle center *(xc(i,j),0)* is encountered) we need to consider the points located at a distance larger than *R(i,j)* from *xc(i,j)* in the distance-based ordering. We can binary search the maximum position *pos* ($0 \leq pos \leq N$) in the ordering such that all the points between the positions *1* and *pos* in the distance ordering are at a distance at most equal to *R(i,j)* from *xc(i,j)*. Then, among all the points located on positions *pos+1, ..., M* (where *M* is the current number of type *2* events processed) we need to find the point with the largest *x* coordinate smaller than *x(i)* (i.e. its index in the original ordering of the *x* coordinates is the largest one smaller than *i*). Let this index be *idx*. Let *idx2* be the smallest index of a point on one of the positions *pos+1, ..., M*, such that *idx2>i*. Either *idx*, *idx2* or both may not exist. If both *idx* and *idx2* exist then we set *pleft(i,j)=idx+1* and *pright(i,j)=idx2-1*. If only *idx* exists and *pleft(i,j)* is undefined then we set *pleft(i,j)=idx+1*. If only *idx2* exists and *pright(i,j)* is undefined then we set *pright(i,j)=idx2-1*. During the sweep we only need to maintain the balanced trees *bt(i)*, containing the *x* coordinates (or indices) of the points located on the positions *i, ..., M* ($1 \leq i \leq M$). The total size of these $O(N)$ trees is $O(N^2)$.

When a type *1* event occurs, then the adjacent points *(x(i),y(i))* and *(x(j),y(j))* are swapped in the distance-based ordering (if they are not adjacent anymore at the moment the event occurs, then the event is simply ignored). Let's assume that they were located on the positions *pos* and *pos+1*. We will update *bt(pos)* and *bt(pos+1)* accordingly (we remove the point which was previously in the balanced tree and we insert the other). In order to identify type *1* events efficiently, we will maintain a *swap events* heap (this way, we will always be able to extract the next event, i.e. the one with the smallest *x* coordinate). Then, we will add to the swap events heap the swap events corresponding to the swap of the points *(x(i),y(i))* and *(x(j),y(j))* with their new neighbors. When a new point *(x(i),y(i))* is encountered (type *2* event), this point is inserted into its appropriate position in the distance-based ordering in *O(N)* time (considering that *xs=x(i)*). The trees *bt(\*)* can be updated with *O(N)* operations. First, we rename the trees corresponding to positions larger than the position *pos* on which *(x(i),y(i))* is inserted in the ordering. Then, we construct *bt(pos)* from scratch and we insert *x(i)* (or just *i*) in the trees *bt(pos'<pos)*. Only *O(1)* swap events (corresponding to the swap of the newly inserted point with its at most two neighbors in the ordering) are inserted into the swap events heap. Let's assume that two points *(x(i),y(i))* and *(x(j),y(j))* are located on consecutive positions in the distance-based ordering (with the first point located on a smaller position) and let's assume that the current x coordinate of the sweep line is *xs*. Then, there exist some points *(xswap,0)* when the order of the two points in the distance ordering needs to be swapped. These points can be computed as: $(x(j)-xswap)^p+y(j)^p=(x(i)-xswap)^p+y(i)^p$. This is a polynomial with degree *p-1* and variable *xswap*. The roots of this polynomial correspond to possible values of *xswap*. We will insert events in the swap events heap for each real value of *xswap* which is larger than the current value of *xs*.

We will proceed in a symmetric manner in order to sweep the plane from right to left. When type *3* events occur, we will be interested in the same two indices *idx* and *idx2* and we will execute the same actions. In the end we will consider all the pairs *(i,j)* again ($1 \leq i \leq j \leq N$). If *pleft(i,j)* is undefined then we will set it to *1*; if *pright(i,j)* is undefined then we will set it to *N*.

There are $O(N^2)$ type *1* events, each of which can be handled in $O(log(N))$ time (due to the swap events heap). There are $O(N)$ type *2* events, each of which can be handled in $O(N \cdot log(N))$ time. There are $O(N^2)$ type *3* events, each of which can be handled in $O(log(N))$ time. The overall time complexity of this algorithm is $O(N^2 \cdot log(N))$.

## 4. RELATED WORK

A divide-and-conquer algorithm for computing the lower envelope of univariate functions on a given interval, where any two functions may intersect only in a bounded number of points, was presented in [1]. This algorithm provided the source of inspiration for one of our solutions to the largest empty circle problem. The problem of placing *K* centers on a line has been studied in [2] (considering the $L_2$ norm). An algorithm with a worse time complexity of $O(N^4 \cdot log(N))$ was given there, compared to the solution proposed by us. Geometric approaches were used in [3] in order to compute multimedia delivery bandwidth-buffer tradeoff functions in the case of multiple servers. The work presented in this paper continues the work presented in [4].

## 5. CONCLUSIONS AND FUTURE WORK

In this paper we presented a set of new server placement algorithms for some constrained geometric problems. Some of the presented algorithms are better than previously known solutions for the same (or similar) problems. In order to be truly useful in practical settings, the presented algorithms should be combined with other, orthogonal, techniques (e.g. risk assessment methods [5-7]).

## 6. ACKNOWLEDGEMENTS

The work presented in this paper was funded by CNCS-UEFISCDI under research grants PD_240/2010 (AATOMMS - contract no. 33/28.07.2010) and ID_1679/2008 (contract no. 736/2009), and by the Sectoral Operational Programme Human Resources Development 2007-2013 of the Romanian Ministry of Labour, Family and Social Protection through the financial agreements POSDRU/89/1.5/S/62557 and POSDRU/6/1.5/S/16.